\documentclass{aa}

\usepackage{graphicx}
\usepackage{txfonts}
\usepackage[hidelinks]{hyperref}
\usepackage{enumitem}
\usepackage{caption}
\usepackage{subcaption}
\usepackage{placeins}
\usepackage{amsmath}
\usepackage{cleveref}
\usepackage{lineno}
\usepackage{amssymb}
\usepackage{xcolor}

\usepackage{orcidlink}

\usepackage{xcolor}   

\newcommand{\rev}[1]{#1}
\newcommand{\revii}[1]{#1}

\begin{document}

   \title{Analysing Turbulent Energy Cascade in a Coronal Mass Ejection using Empirical Mode Decomposition}


   \author{Akanksha Dagore
          \inst{1,2}\orcidlink{0009-0001-5407-3016}, 
          Giuseppe Prete \inst{2}\orcidlink{0000-0003-3739-3170}, 
          Vincenzo Capparelli \inst{2}\orcidlink{0009-0005-0463-8612}, 
          Vincenzo Carbone \inst{2}\thanks{Deceased January 21, 2025}   \orcidlink{0000-0002-3182-6679},
          Fabio Lepreti \inst{2}\orcidlink{0000-0001-5196-2013}
          }

   \institute{Dipartimento di Fisica, Università di Trento, via Sommarive 14, 38123 Povo (TN), Italy \\
        \email{akanksha.dagore@unitn.it}
   \and
    Dipartimento di Fisica, Università della Calabria, via P. Bucci - Cubo 31/C, 87036 Rende (CS), Italy \\
             }

   \date{Received September 24, 2025; accepted -}

 
  \abstract
   {Coronal mass ejections (CMEs) are large-scale expulsions of plasma and magnetic flux that emanate from the Sun's corona and travel into the heliosphere. Once in interplanetary space, they are referred to as interplanetary coronal mass ejections (ICMEs), \rev{often characterised} by a shock wave, a sheath region, and, in some cases, a magnetic cloud. ICMEs are capable of driving magnetic disturbances that can trigger geomagnetic storms and significantly influence the space weather.}
   {The turbulent nature of CMEs has been well established by previous studies using Fourier and wavelet-based methods. In the present work, we apply empirical mode decomposition (EMD) in conjunction with Hilbert spectral analysis (HSA) to investigate the turbulence characteristics at different stages of the ICME event observed on 27 June 2013 by the MAG instrument onboard NASA's ACE spacecraft.}
   {The ICME event is divided into \rev{four regions}: (i) the preceding solar wind, (ii) the sheath region, (iii) the magnetic cloud, and (iv) the trailing solar wind. The magnetic field components ($B_{x}$, $B_{y}$, and $B_{z}$) of \rev{the four regions} are decomposed into intrinsic mode functions (IMFs) using EMD. The instantaneous frequencies and amplitudes of the extracted IMFs are derived using HSA to generate the Hilbert spectra. The energy distribution across frequencies is computed from the second-order marginal Hilbert spectra, and the spectral slopes in the inertial range are calculated from linear fits.}
   {For the solar wind preceding the ICME shock, the Hilbert spectral analysis shows a slope close to the Kolmogorov value ($\alpha_{HHT}\approx -1.68$), indicating a fully developed turbulence at 1 AU. A clear steepening is observed in the sheath and in the trailing solar wind ($\alpha_{HHT}\approx -1.78$ and $-1.79$), consistent with enhanced intermittency and intensified non-linear activity associated with shock compression and solar wind–ICME interactions. Within the magnetic cloud, the scaling exponent becomes slightly less steep ($\alpha_{HHT}\approx -1.71$), \revii{suggesting that the effects causing the spectral steepening in the sheath and downstream wind are less prevalent inside the large-scale flux ropes.}}
   {\rev{Our results show that ICME passage modifies the turbulent energy distribution across scales. The EMD-HSA method provides smoother and more stable spectral estimates than the conventional Fourier approach, making it well suited for analysing turbulence in non-stationary solar wind structures.}}


   \authorrunning{A. Dagore et al}
   \maketitle
%

\section{Introduction}

Coronal mass ejections (CMEs) are eruptions of tremendous amounts of plasma and magnetic flux from the Sun's corona that propagate into the heliosphere. The matter expelled comprises primarily charged particles in the form of protons and electrons that are \rev{carried with the solar wind} towards interplanetary space. These particles are embedded in the ejected magnetic field along twisted helical coherent structures, known as flux ropes, which result from the magnetic reconnection of field lines in the Sun's corona \citep{chintzoglou2015formation, chen2017physics, 2017SoPh..292...71J, james2018observationally}. The first evidence of a documented CME dates back to 1971 \citep{tousey1973space} which was recorded by the white-light coronagraph onboard Orbiting Solar Observatory 7 (OSO 7) \citep{koomen1975white}, and since then, CMEs have continued to garner significant interest of researchers in the field of solar physics.

Previously conducted studies have attempted to classify CMEs based on their structural and intrinsic properties. \cite{howard1985coronal} studied 998 CMEs observed from 1979 to 1981 and categorised them into nine major classes based on their morphological characteristics, thereby suggesting a strong dependence of the CME properties on its structure. \cite{nicewicz2016classification} conducted a quantitative analysis of 6621 CMEs recorded between 1996 and 2004, and proposed four categories based on their acceleration, angular width, velocity, and mass. In 2013, the Space Weather Research Center (SWRC) at NASA's Goddard Space Flight Center (GSFC) introduced the CME SCORE scale, which categorises CMEs into five classes according to their speeds, ranging from less than 500 km/s to over 3000 km/s \citep{evans2013score}. However, no previous studies have been performed that can reliably classify CMEs \rev{on their turbulence properties from remote imaging.}

CMEs travelling farther away from the Sun and entering interplanetary space are referred to as interplanetary coronal mass ejections (ICMEs) \citep{gosling1991geomagnetic, kilpua2017geoeffective, song2020all, zhang2021earth, davies2021situ}.
\textit{In situ} observations have shown ICMEs \rev{to often exhibit a forward shock} that accelerates the charged particles from the solar ejecta towards interplanetary space. These charged particles are capable of triggering disturbances in the magnetosphere, resulting in magnetic storms and indicating the arrival of a possible ICME. The passage of the ICME-driven shocks is followed by a region of enhanced magnetic field, known as the ICME sheath, which forms due to the combined mechanism of CME propagation and expansion \rev{\citep{siscoe2008ways}}. It is characterised by an abrupt increase in the solar wind temperature, density, and speed, making it an extensive laboratory to study the plasma properties of the solar wind \citep{kilpua2017coronal, DVN/C2MHTH_2024}.
An ICME-driven shock is usually accompanied by a magnetic cloud (MC), a term introduced by \rev{\cite{burlaga1981magnetic}} to address the gas clouds emitted from the Sun into the interplanetary medium. An MC can be distinguished from its surrounding solar wind by enhanced magnetic field, smoothly rotating magnetic field components due to the presence of flux ropes, reduced plasma beta ratio, i.e. plasma pressure to magnetic pressure ($\beta<1$), and weaker proton temperature and density resulting from its expansion into the interplanetary space \citep{burlaga1981magnetic, klein1982interplanetary, gosling1987field, rouillard2011relating}.
ICMEs play a crucial role in influencing space weather and are responsible for the origin of some of the most intense geomagnetic storms \citep{wu2002effects, wu2006relationships}.

Investigations have shown that ICMEs display \rev{enhanced and more intermittent turbulence} in the sheath than in the preceding ambient solar wind \citep{sorriso2021turbulent, kilpua2021statistical, ghuge2025turbulent}. This can be attributed to the compressed nature of plasma in the sheath, resulting from the power injected earlier by the shock. Moreover, the solar wind itself is known to exhibit turbulent, anisotropic, and intermittent behaviour \citep{bruno2013solar}, resulting in energy-transfer processes that govern plasma heating and particle acceleration. Early evidence of turbulence in the solar wind was reported by \cite{1968ApJ...153..371C} who found the presence of turbulent fluctuations in the magnetic field and the radial velocity component by analysing their spectral densities using data from the Mariner 2 spacecraft at 1 AU. Since then, studies have shown that cross-scale energy transfer due to turbulence in the solar wind follows a power-law behaviour across the injection, inertial, and dissipation ranges. For fully developed turbulence observed at 1 AU via in situ measurements, it was found that the power-law scaling resembles that of the hydrodynamic turbulence, as described by \cite{1941DoSSR..30..301K}. 

Turbulence in solar wind and CMEs have been studied extensively using some of the well-established methods of Fourier analysis and Morlet wavelet analysis \rev{\citep{zhao2021turbulence, soljento2023imbalanced, ruohotie2025intermittency}}. In this work, however, we employ the novel technique of empirical mode decomposition (EMD) \citep{huang1998empirical} to analyse the turbulence properties across different regions of an ICME event. This method breaks down the signal into a finite set of simpler modes. The Hilbert transform is applied to each of the obtained modes to extract its instantaneous frequency and amplitude, a process known as Hilbert spectral analysis (HSA). While the Fourier and wavelet-based methods assume local stationarity and linearity in the signal data, the EMD-HSA framework is not restricted to these constraints and is shown to work well with non-linear, non-stationary, and turbulent data, which are some of the common characteristics of solar wind signals. The technique has previously been applied to solar wind data to study the scaling properties and identify the spectral exponents associated with turbulence \citep{carbone2018arbitrary, alberti2019multifractal, carbone2020statistical, carbone2021statistical}, examine the temporal behaviour of solar magnetic field at different heliographic latitudes \citep{vecchio2012dynamics}, derive rotational cycle and stability of solar mean magnetic flux elements \citep{xiang2016ensemble}, and analyse the long-term solar variability and grand minima occurrence \citep{vecchio2017connection}. However, to the best of our knowledge, EMD-HSA methodology has not been used to analyse the turbulence characteristics across different ICME stages.

EMD has found widespread application in many fields that deal with non-linear and non-stationary real time-series data. These include a broad spectrum of scientific and applied domains such as seismology \citep{battista2007application, raghukanth2012empirical, han2013empirical}, marine sciences \citep{schmitt2009analysis, ezer2013gulf, huang2014time}, meteorology \citep{coughlin2005empirical, molla2006empirical, capparelli2013spatiotemporal, martins2017using}, engineering \citep{loutridis2004damage, xu2004structural, lei2013review}, and medicine \citep{echeverria2001application, liang2005application}, as well as strategic planning domains such as economics and finance \citep{ zhang2008new, zhu2015carbon}.

In this study, we present our findings on the turbulence power spectra derived using the EMD-HSA framework applied to magnetic field data ($B_x$, $B_y$, and $B_z$) measured in the Geocentric Solar Ecliptic (GSE) coordinates across different regions of an ICME event. Section 2 outlines the characteristics of the ICME selected for this analysis, along with a description of the data. The EMD-HSA methodology is explained in Section 3. The results of the analysis are presented in Section 4, followed by the conclusions reported in Section 5.


\section{Data}

In this work, we studied the ICME event observed on 27 June 2013 at \rev{13:30} UTC by NASA's Advanced Composition Explorer (ACE) spacecraft located at the Sun-Earth's L1 Lagrange point. \rev{The event has also been previously analysed by \cite{prete2024euhforia} using the EUHFORIA simulation model.} Operating in a Lissajous orbit, the primary mission of the ACE spacecraft is to study energetic particles originating from the Sun, which can contribute to our understanding of the solar system and the Milky Way galaxy. \rev{Additionally, ACE was used to continuously monitor space weather conditions near the L1 point, until its replacement by DSCOVR in 2015 \citep{burt2012deep}}.

\begin{figure*}
    \centering
    \includegraphics[width=\linewidth]{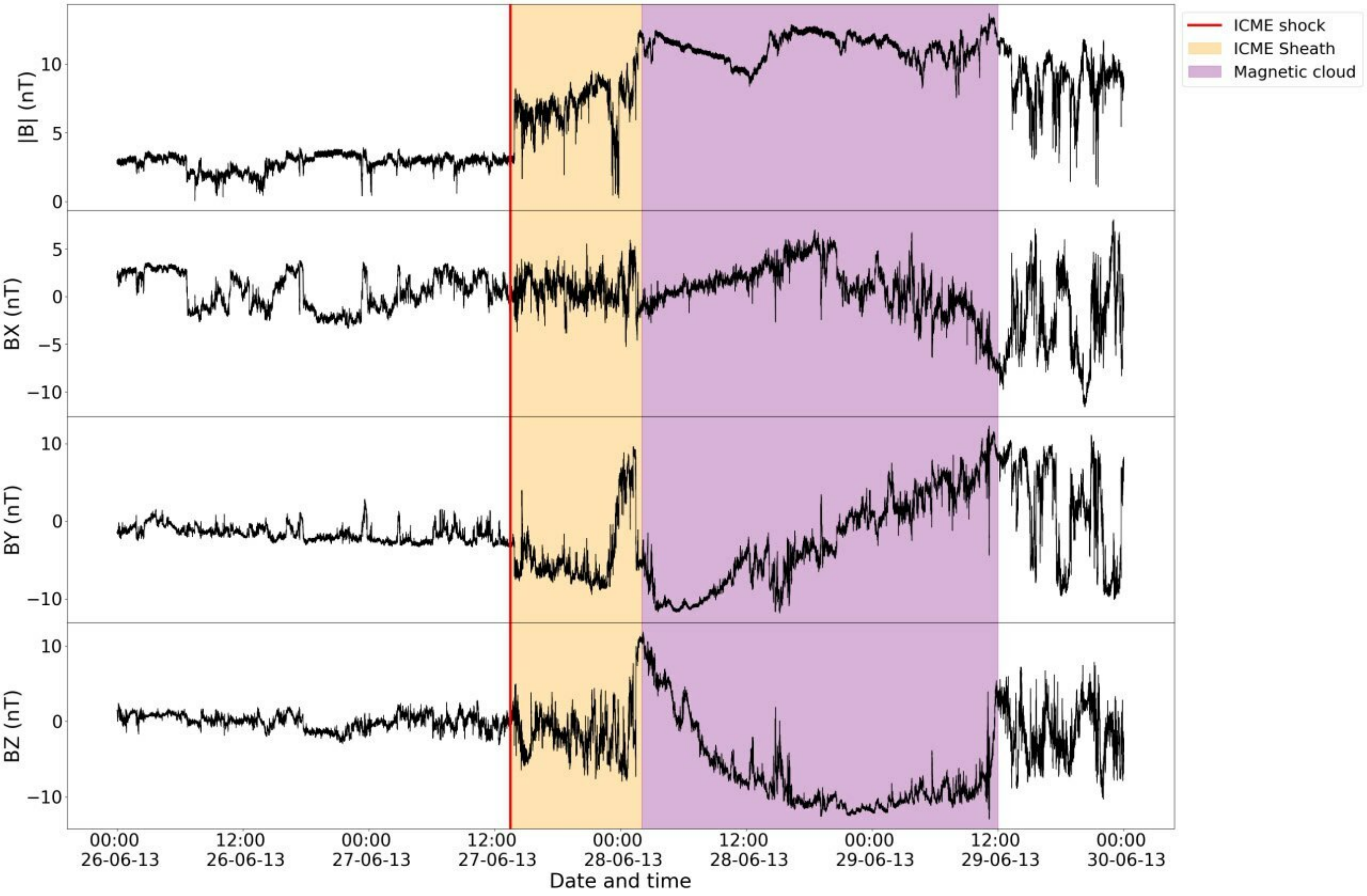}
    \caption{Magnetic field data recorded by NASA's MAG instrument onboard ACE. The subplots represent the signal for the field magnitude $|B|$ and the component vectors $B_{x}$, $B_{y}$, and $B_{z}$ in the GSE coordinate system. The red vertical line denotes the shock arrival resulting in the ICME sheath indicated by the orange-shaded region. This is followed by the MC, shown as violet-shaded region. The onset of the sheath and the end of the MC spans the entire ICME event.}
    \label{CME_B_field}
\end{figure*}

The observational data for the ICME under investigation was collected by the Magnetometer (MAG) instrument onboard the ACE spacecraft \citep{chiu1998ace}. The ICME was preceded by a fast forward shock and featured an embedded MC \citep{burlaga1981magnetic}, characterized by the dominance of magnetic pressure over plasma pressure, and exhibiting a rotating magnetic field with signatures consistent with flux rope structures and field rotations between 90°- 180° \citep{2018SoPh..293...25N}. The World Data Center for Geomagnetism, Kyoto\footnote{\url{https://wdc.kugi.kyoto-u.ac.jp/index.html}} recorded a minimum geomagnetic equatorial Dst index of -101 nT during the MC passage, indicating a geomagnetic storm that can be classified as `strong' \citep{gonzalez1994geomagnetic, loewe1997classification}. A commonly used indicator to measure geomagnetic storm intensity, the Dst index is defined as the drop in the Earth's horizontal magnetic field component resulting from ring current enhancements.

The data products were downloaded in the Level 2 format with a time resolution \rev{$\Delta t$} = 1 second, comprising the magnetic field measurements in the GSE coordinate system. The downloaded data spanned a four-day period, from 26 June 2013 at 00:00:00 UTC to 29 June 2013 at 23:59:59 UTC, capturing the key features of the ICME event --- the shock, sheath, and MC region. The ICME was characterised by the arrival of a shock on 27 June 2013 at \rev{13:30 UTC\footnote{\url{http://www.ssg.sr.unh.edu/mag/ace/ACElists/obs_list.html}}} and the termination of the MC on 29 June 2013 at 12:00 UTC. The sheath spanned from 27 June 2013, \rev{13:30} UTC to 28 June 2013, 02:00 UTC, and the magnetic cloud extended from 28 June 2013, 02:00 UTC to 29 June 2013, 12:00 UTC \citep{DVN/C2MHTH_2024}. Fig. \ref{CME_B_field} represents the total magnetic field magnitude |$B$| and the magnetic field components ($B_{x}$, $B_{y}$, $B_{z}$) in GSE coordinates for the different ICME stages. The red vertical line marks the arrival of the ICME shock. This is followed by the ICME sheath, which is colour-coded in orange, and later by the MC, as shown by the violet-shaded region. The \rev{mean field magnitude ($B$)} and the mean solar wind velocity in the ICME ($V_{SW}$) were reported to be 10 nT and 390 km/s, respectively.
 
\rev{The turbulence characteristics were analysed separately for each region and component. The approximate duration of each region is presented in Table \ref{table:icme_duration}.}

\begin{table}
\caption{\rev{Approximate duration (in hours) for each region of the ICME event.}}
\label{table:icme_duration}
\centering
\begin{tabular}{c c}
\hline\hline
ICME region & Duration (hours) \\ 
\hline
Preceding trailing solar wind & 37.5 \\
Sheath & 12.5 \\
MC & 34.0 \\
Trailing solar wind & 12.0 \\
\hline
\end{tabular}
\end{table}


\section{Methodology}

In this study, we applied the methods of EMD and HSA on the magnetic field component signals $B_{x}$, $B_{y}$, and $B_{z}$ to study the nature of turbulence across the different ICME \rev{regions} defined above. The EMD method decomposes the original signal into simpler components, each representing a distinct oscillatory mode, while the HSA method extracts the instantaneous quantities from these modes. The combined approach of EMD and HSA is designated by NASA as the Hilbert-Huang transform (HHT) \citep{kizhner2004hilbert, huang2005empirical, huang2008review, doi:10.1142/S179353691203001X}. This section offers a detailed overview of these techniques and their application to our data.

\subsection{Empirical Mode Decomposition}

The empirical mode decomposition (EMD) is a robust, data-driven technique for analysing complex signals by decomposing them into simplified components known as intrinsic mode functions (IMFs). It is both adaptive and efficient for processing non-linear and non-stationary time-series, providing insights into the intrinsic characteristics of the data. Each IMF represents a characteristic time scale and frequency that captures a principle mode of oscillation embedded in the signal. A component signal can be classified as an IMF if it satisfies the following two conditions:
\begin{enumerate}[label=(\roman*), wide, labelwidth=!,itemindent=!,labelindent=0pt, leftmargin=0em, noitemsep]
    \item The total number of local extrema (minima and maxima) must be equal to or one greater than the zero-crossings present in the component signal.
    \item The mean value calculated from the upper and lower envelopes, constructed from the local maxima and minima, respectively, must be zero at every point along the signal.
\end{enumerate}

The aforementioned criteria ensure that the resulting IMF is symmetric with respect to the zero mean and eliminate any unwanted large fluctuations in the instantaneous frequencies. The IMFs are extracted from the original signal through an iterative procedure, known as the sifting process, which is described in detail in the following subsection.

\subsubsection{Sifting Process}

The EMD method operates by iteratively identifying and extracting the dominant oscillatory modes present in a time-series signal. In this work, the ICME data of the magnetic field components ($B_{x}$, $B_{y}$, and $B_{z}$) measured across the \rev{four ICME regions} served as the primary signals. We denote this by a generic term $B(t)$. The local extrema, determined within $B(t)$, were used to construct an upper envelope, $\varepsilon_{up}(t)$ and a lower envelope, $\varepsilon_{down}(t)$ around $B(t)$ using spline interpolation. 
The mean of these two envelopes $\varepsilon_{mean}(t)$ was calculated and subtracted from the original signal $B(t)$ to derive the first component signal,
\begin{equation}
    f_{1}(t) = B(t) - \varepsilon_{mean}(t) \ .
\end{equation}
The function $f_{1}(t)$ is considered an IMF if it satisfies the conditions outlined earlier. If the criteria are not met, then a new mean envelope $\varepsilon_{mean 1}(t)$ is computed for $f_{1}(t)$ and subtracted from it to yield $f_{1(1)}(t)$,
\begin{equation}
    f_{1(1)}(t) = f_{1}(t) - \varepsilon_{mean1}(t) \ .
\end{equation}
This sifting process is repeated $p$ times until the function $f_{1(p)}(t)$ satisfies the criteria to be classified as an IMF. Once this condition is met, the resulting function is identified as the first IMF, denoted by $C_{1}(t)$. This quantity is then subtracted from the original signal to compute the first residual of the data,
\begin{equation}
    r_{1}(t) = B(t) - C_{1}(t) \ .
\end{equation}
The residual $r_{1}(t)$ then serves as the new signal on which the sifting process is repeated to extract the next IMF.

The objective of the sifting process is to isolate oscillatory modes by smoothing out irregular amplitude fluctuations. However, excessive sifting could also result in loss of crucial signal information. Therefore, to preserve the important signal components, the process was controlled by implementing a stopping criterion. In this work, we adopted the stopping criterion 
proposed by \cite{huang1998empirical}, which introduced a standard deviation term, $\sigma$ between two consecutive siftings such that,
\begin{equation}
    \sigma = \sum_{t=0}^{T}\left(\frac{|f_{1(p)}(t) \ - \ f_{1(p-1)}(t)|^{\revii{2}}}{f_{1(p-1)}^{2}(t)}\right) \ ,
\end{equation}
where, $T$ denotes the total duration of the signal. For our dataset, a threshold of $\sigma=0.1$ was selected to terminate the sifting process, ensuring a balance between retaining signal features and avoiding excessive iterations.

\begin{figure*}
\centering
\makebox[\textwidth][c]{%
\begin{subfigure}{0.54\textwidth}
    \centering
    \includegraphics[width=1.005\textwidth, height=1.18\textwidth]{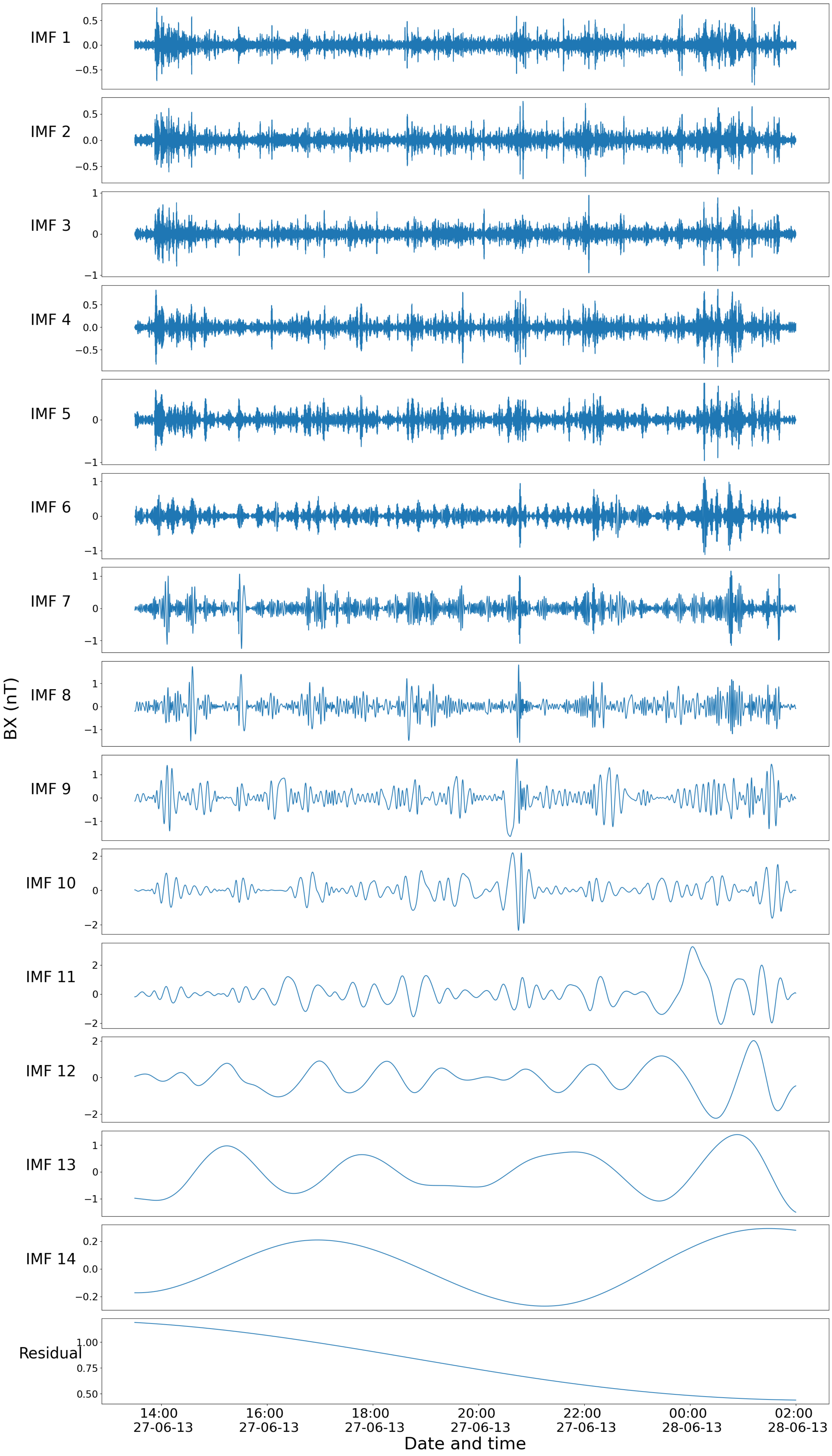}
    \label{fig:Bx}
\end{subfigure}
\begin{subfigure}{0.54\textwidth}
    \centering
    \includegraphics[width=1.005\textwidth, height=1.18\textwidth]{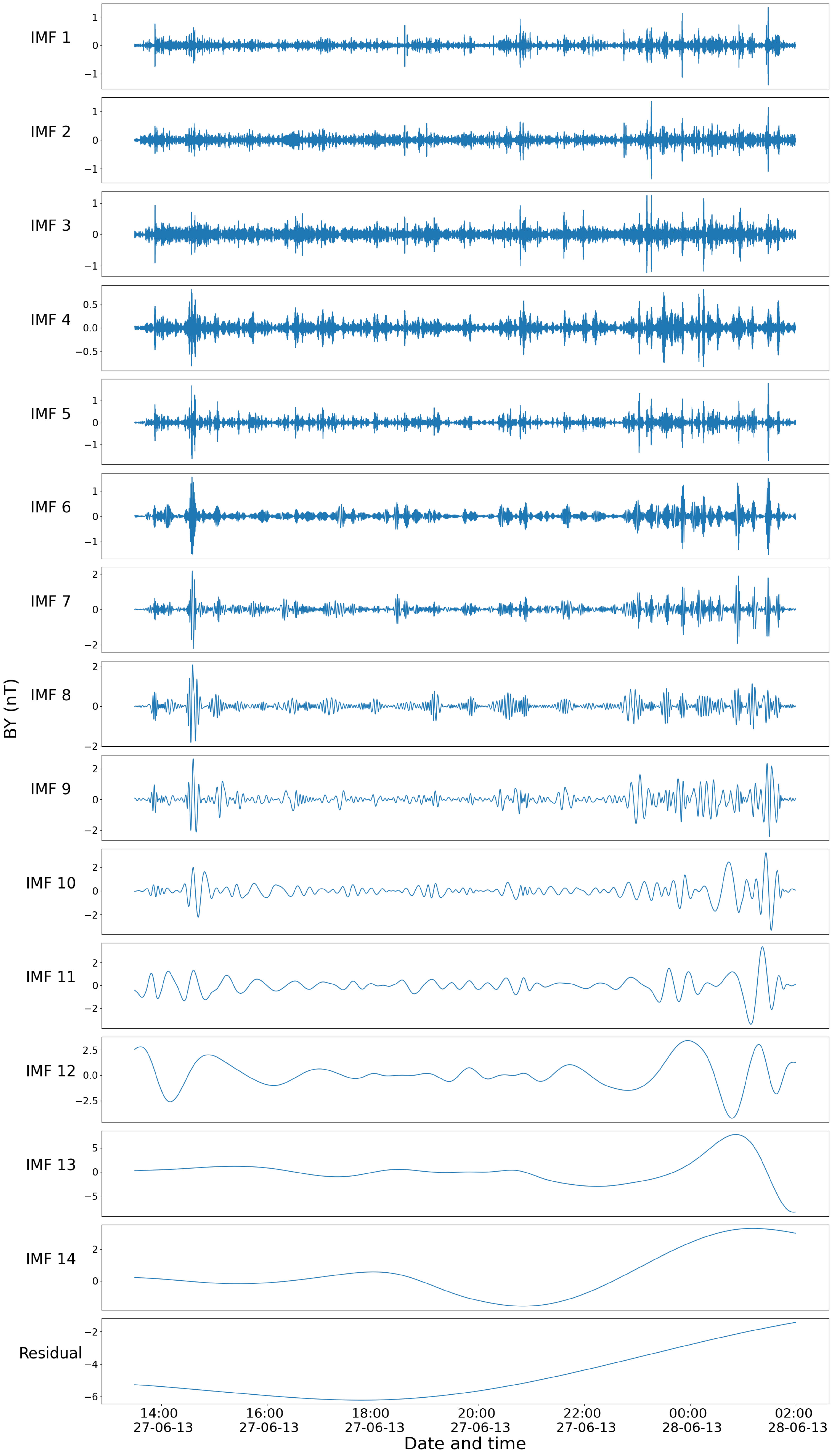}
    \label{fig:By}
\end{subfigure}}
\makebox[\textwidth][c]{%
\begin{subfigure}{0.54\textwidth}
    \centering
    \includegraphics[width=1.005\textwidth, height=1.18\textwidth]{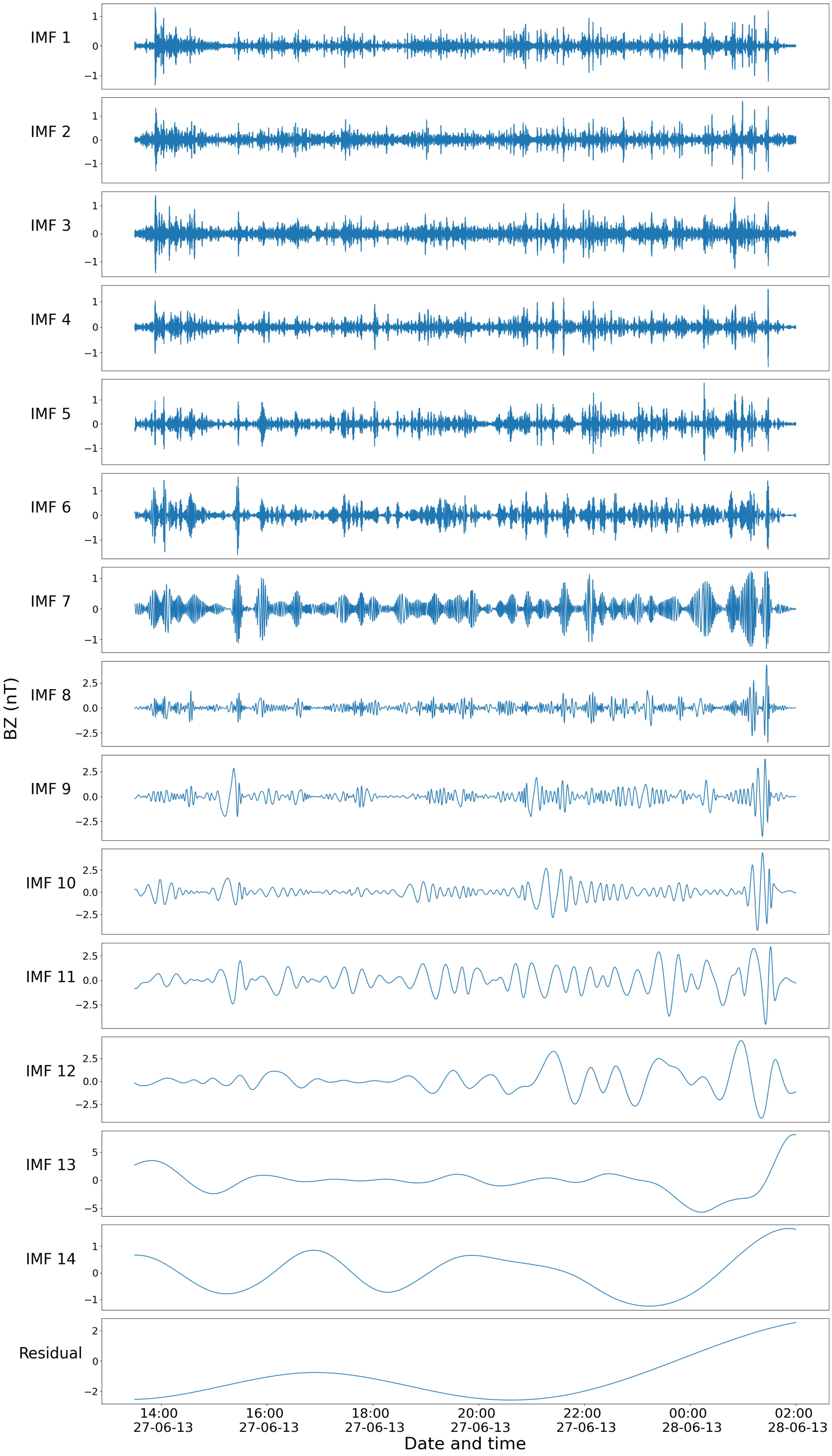}
    \label{fig:Bz}
\end{subfigure}}
\vspace{-0.9cm}
\caption{IMFs obtained for the magnetic field components $B_{x}$, $B_{y}$, and $B_{z}$ \rev{for the \revii{ICME} sheath}. The y-axis represents the magnitude (nT) and the x-axis denotes the time duration of the interval.}
\label{IMFs}
\end{figure*}

For a signal containing $N$ data points, the approximate number of IMFs that can be extracted is $s \approx log_{2}N$. The first IMF captures the highest frequency components of the signal, typically corresponding to the smallest time scales and often dominated by the noise present in the signal. The subsequent IMFs represent progressively smaller frequencies (larger time scales), as the higher frequency components get filtered out in the previous IMFs. Consequently, the IMFs obtained from EMD are inherently ordered from high to low frequency. \rev{The sifting process was repeated until a monotonic function or a function from which no further IMFs can be extracted was reached. We denote this as the final residue, $r(t)$.}
The original signal $B(t)$ can then be reconstructed by summing all the IMFs and the residual,
\begin{equation}
    B(t) = \sum_{j=1}^{s}C_{j}(t) \ + \ r(t) \ .
\end{equation}
The extracted IMFs from $B_{x}$, $B_{y}$, and $B_{z}$ for the \revii{ICME} sheath region are shown in subplots (a), (b), and (c) respectively of Fig. \ref{IMFs}. \rev{The IMFs depict enhanced fluctuations, capturing the heightened magnetic activity following the shock.}

\subsection{Hilbert Spectral Analysis}

Following the extraction of IMFs from the component magnetic field signals, Hilbert spectral analysis (HSA) was performed on each of the resultant IMFs to extract its instantaneous quantities. The procedure involves applying the Hilbert transform to each IMF,
\begin{equation}
    \mathcal{H}\{C_{j}(t)\} = \frac{p}{\pi}\int_{-\infty}^{+\infty}\frac{C_{j}(\tau)}{t-\tau}\ d\tau \ ,
\end{equation}
where $p$ denotes the Cauchy principal value. The application of the Hilbert transform results in a real-valued signal having a phase shift of $\pm90^{\circ}$ in the frequency domain, allowing the construction of a complex analytic signal. By combining $C_{j}(t)$ and $\mathcal{H}\{C_{j}(t)\}$, the analytic signal $z(t)$ can be written as follows,
\begin{equation}
    z(t) = C_{j}(t) \ + \ i\mathcal{H}\{C_{j}(t)\} = a_{j}(t)e^{i\phi_{j}(t)} \ .
    \label{analytic_signal}
\end{equation}
Here, $a_{j}(t)$ and $\phi_{j}(t)$ represent the instantaneous amplitude and instantaneous phase, respectively of the $j^{th}$ IMF, and are defined as,
\begin{align}
    a_{j}(t) & = \sqrt{(C_{j}(t))^{2} + (\mathcal{H}\{C_{j}(t)\})^{2}} \ , \\
    \phi_{j}(t) & = \tan^{-1}\left(\frac{\mathcal{H}\{C_{j}(t)\}}{C_{j}(t)}\right) \ .
\end{align}
The instantaneous frequency was then computed as,
\begin{equation}
    \omega_{j}(t) = \frac{d\phi_{j}(t)}{dt} \ .
\end{equation}
Using Eq. \ref{analytic_signal}, the original signal can then be reconstructed by summing the real parts of the analytic signals corresponding to each IMF,
\begin{equation}
    B(t) = \mathrm{Re}\sum_{j=1}^{s}a_{j}(t)e^{i\int\omega_{j}(t)dt} \ .
\end{equation}
Here, $r(t)$ was dropped as it represented a {constant or monotonic function.} The measure of amplitude across the frequency-time domain yields the Hilbert spectrum, $H(\omega,t)$. The total energy contribution from each frequency can be obtained by integrating $H(\omega,t)$ over the complete time interval $T$, resulting in the marginal Hilbert spectra,
\begin{equation}
    h(\omega) = \frac{1}{T}\int_{0}^{T}H(\omega,t)\ dt \ . \label{eq:marginal_hht}
\end{equation}


\section{Analysis and Results}

In this section, we report our findings regarding the energy contributions associated with different frequencies present in the magnetic field of the selected ICME. The analysis includes all three magnetic field components, $B_{x}$, $B_{y}$, and $B_{z}$ across the \rev{four ICME regions}. For each case, the EMD-HSA method was applied to construct the Hilbert spectrum, providing the amplitude distribution over the time-frequency domain. This was followed by the computation of the second-order marginal Hilbert spectrum for each \rev{region}. A linear best-fit analysis was performed on the second-order marginal spectra and the resulting slopes were compared with the theoretical Kolmogorov scaling. The detailed results and interpretations are discussed in the following subsections.

\subsection{Hilbert Spectra}
\label{subsec_hilbert_spectra}

\begin{figure*}
    \centering
    \includegraphics[width=\linewidth]{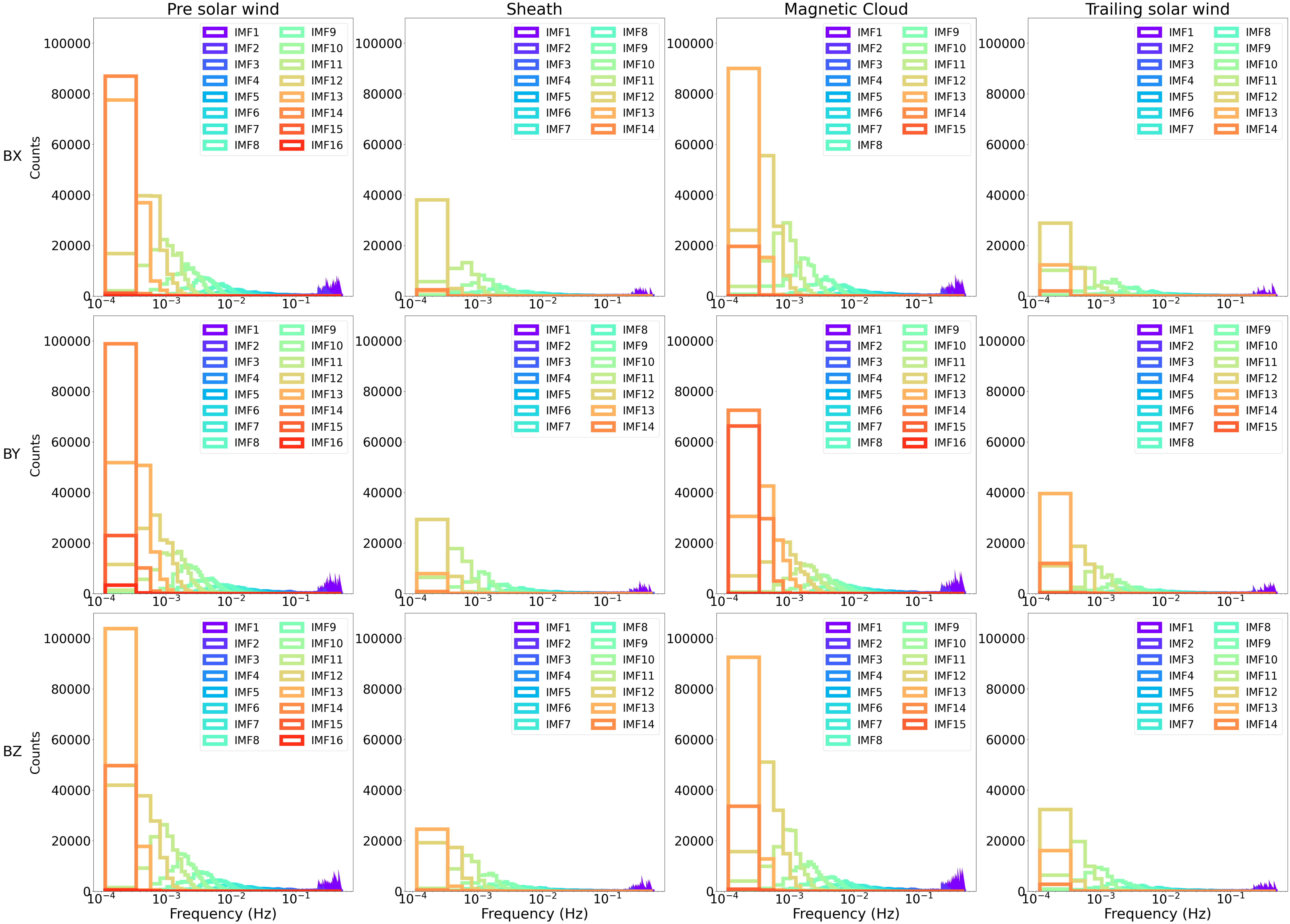}
    \caption{Instantaneous frequency distribution for each IMF for all the magnetic field components ($B_{x}$, $B_{y}$ and $B_{z}$), \rev{across different ICME regions}. The instantaneous frequencies are binned into \rev{2160} equally spaced logarithmic bins ranging from 0 to \rev{0.5} Hz. Each IMF order is represented by a distinct colour to enhance clarity and differentiation.}
    \label{freq_dist}
\end{figure*}

The Hilbert transform was applied to all the extracted IMFs from the magnetic field components across the \rev{ICME regions} to compute their corresponding instantaneous amplitudes and frequencies. Fig. \ref{freq_dist} shows the distribution of the instantaneous frequencies for each IMF across all magnetic field components and \rev{ICME regions}. The instantaneous frequency distribution was grouped into \rev{2160} equal-width logarithmic bins, spanning from 0 to \rev{0.5} Hz. 

To construct the Hilbert spectrum, the range of instantaneous frequencies was divided into \rev{2160} equally spaced bins to resolve fine-scale features in the signal. Unlike Fourier spectral analysis, where frequency range is determined by the total duration of the signal ($T$) and its sampling frequency ($\Delta \omega$), the instantaneous frequencies ($\omega_{j}$) obtained by the HSA method represent the occurrence of specific frequencies at a local time instance $t$. Consequently, the frequency resolution in HSA can be adjusted by modifying the bin size that stores these instantaneous frequencies. While this flexibility allows examination of fine spectral details, selecting an arbitrary frequency resolution might lead to an erroneous interpretation of the spectrum \citep{huang2011hilbert}. Broad bins may merge distinct frequencies, obscuring fine details, whereas narrower bins improve frequency resolution but increase computational demands, particularly when analysing large datasets.

For a signal sampled at a time resolution $\Delta t$, the highest frequency that can be recorded after applying HSA is given by the Nyquist frequency, $f_{q} = 1/(2\Delta t)$ Hz. \rev{In our case, where the signal duration differs for each ICME region, the finest resolvable frequency structure, $\Delta f = 1/T$ is given by the region with the shortest duration.}
The bin size $b$ to generate the Hilbert spectrum can then be determined by the ratio of the Nyquist frequency to the fundamental frequency resolution. \rev{We scale this with respect to a binning factor $F_{b}$ to prevent noise domination at higher frequencies and to manage computational efficiency,}
\begin{equation}
    b = \frac{f_{q}}{\Delta f} \times \frac{1}{F_{b}} \ .
\end{equation}
To maintain uniform resolution, $F_{b}$ is kept constant across all bins. In our case, \rev{the shortest region corresponds to the trailing solar wind $\approx 12$ hours, yielding $f_{q} = 0.5$ Hz and $\Delta f = 2.32 \times 10^{-5}$ Hz.} Through empirical testing to balance spectral resolution and computational efficiency, we selected \rev{$F_{b} \approx 10$, which gives,}
\begin{equation}
    b = \frac{1/(2\Delta t)}{1/T} \times \frac{1}{ F_{b} } \approx 2160 \ \text{bins} \ .
\end{equation}
Thus, the frequency resolution for each bin becomes, 
\begin{equation}
    \Delta f_{b} = \frac{f_{q}}{b} \approx 2.3 \times 10^{-4} \ \text{Hz} \ .
\end{equation}

\begin{figure*}
    \centering
    \includegraphics[width=\linewidth]{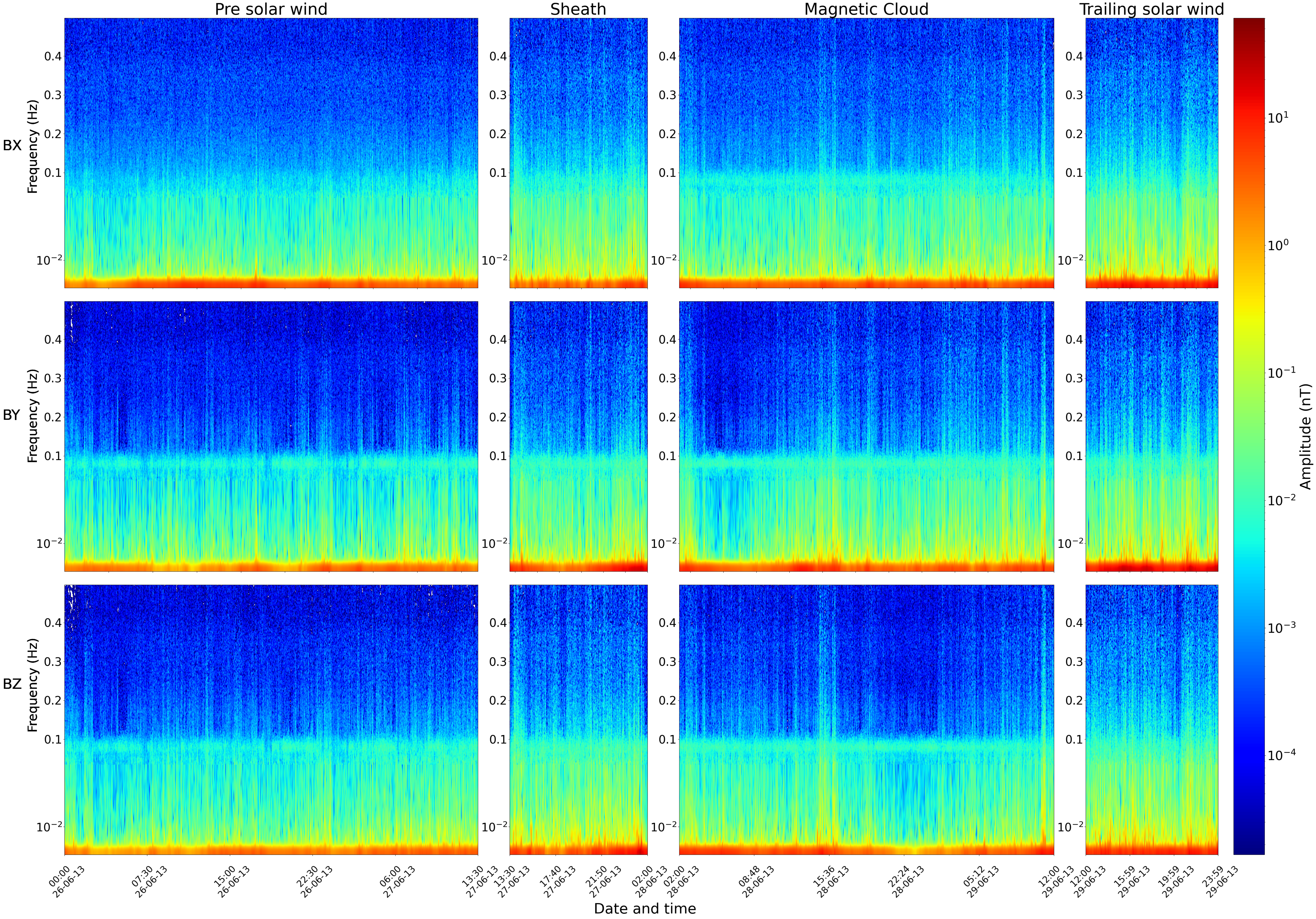}
    \caption{\rev{Hilbert spectra for magnetic field components across different ICME regions. \textit{From left to right}: regions corresponding to the preceding solar wind, sheath, magnetic cloud and trailing solar wind. \textit{From top to bottom:} magnetic field components for $B_{x}$, $B_{y}$, and $B_{z}$. The y-axis is represented on the logarithmic scale up to 0.032 Hz to visually emphasise the contribution of the dominating low frequencies.}}
    \label{hilbert_spectrum}
\end{figure*}

The Hilbert spectrum was generated by summing the amplitude contributions of all IMFs within each frequency bin at every time point. \rev{Fig. \ref{hilbert_spectrum}} illustrate the resulting Hilbert spectra for the ICME event. \rev{The top, middle and bottom panels represent the magnetic field components $B_{x}$, $B_{y}$, and $B_{z}$, while each subplot in the panel correspond to the ICME regions.} A Gaussian filter was applied while generating the heatmaps to produce smoother and visually continuous spectra. The Hilbert spectra indicate that the smaller instantaneous frequencies ($\rev{<0.1}$ Hz) dominate the ICME signal, contributing to the strongest observed amplitudes.

\rev{In the first interval corresponding to the solar wind preceding the ICME shock, weaker amplitudes are observed particularly in the $B_{y}$ and $B_{z}$ components, suggest lower levels of magnetic turbulence, consistent with relatively undisturbed background solar wind conditions. Following the shock, sharp fluctuations in the amplitude levels are observed within the ICME sheath, indicating enhanced magnetic field variations. Being highly compressed plasma structures, the sheath region \revii{displays} strong turbulent fluctuations following the shock passage. The trend continues to be observed within the MC, which is characterised by large-scale magnetic flux ropes. The peaks become sharper and more clearly distinguishable, suggesting the presence of intermittent turbulence. Following the termination of the MC, the trailing solar wind \revii{displays} the strongest amplitudes in all the three magnetic field components of the ICME event, highlighting continued magnetic field variability in the aftermath of the event.} Furthermore, the presence of a constant frequency at $\omega_{sc} \approx$ 0.083 Hz was observed that appeared as a horizontal feature in all the Hilbert spectra. This constant frequency corresponds to the spin rate (gyration frequency) of the ACE spacecraft, which is 5 revolutions per minute (rpm), with the spin axis orientated along the Sun-Earth line \citep{chiu1998ace, stone1998advanced}. 

\begin{figure*}
    \centering
    \includegraphics[width=\linewidth]{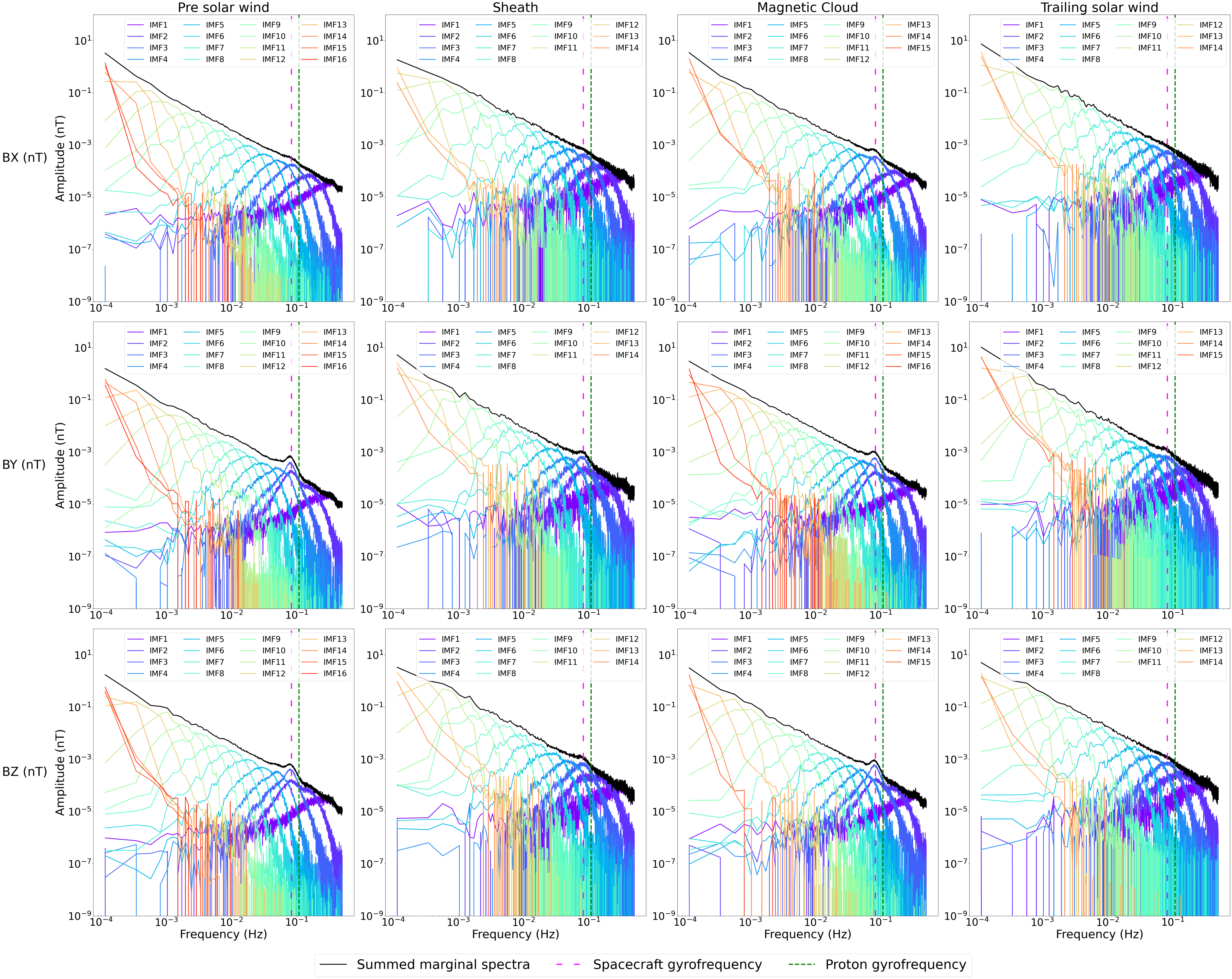}
    \caption{Marginal Hilbert spectra for each IMF representing the total amplitude contribution across the frequency domain. Each IMF order is depicted using a distinct colour for clarity. The black curve in each subplot represents the total marginal spectrum obtained by summing the marginal spectra of each IMF order. The vertical magenta and green dashed lines mark the spacecraft gyrofrequency and proton gyrofrequency, respectively.}
    \label{marginal_amp}
\end{figure*}

The marginal Hilbert spectra for the magnetic field components \rev{for each ICME region} were computed using Eq. \ref{eq:marginal_hht} and are shown in Fig. \ref{marginal_amp}. Each subplot depicts the marginal Hilbert spectrum for the extracted IMFs, represented in a distinct colour, while the black curve depicts the total marginal spectrum obtained by summing the spectrum of each IMF. The results of Fig. \ref{marginal_amp} demonstrate the capability of the EMD-HSA method in filtering the signal into finite frequency components. Modes 1 and 2 correspond to time scales shorter than the proton gyrofrequency (marked by the vertical green dashed line) that fall within the turbulence dissipation range. Modes 5 to 12 represent larger time scales and correspond to the inertial range of the turbulence scale. The mode corresponding to $\omega_{sc}$ (indicated by the vertical magenta dashed line) was captured by modes 2, 3, and 4 and appears as a small bump in each subplot. This feature is more enhanced in the marginal spectra of the components $B_{y}$ and $B_{z}$ compared to the component $B_{x}$, \rev{possibly due to the orientation of the spacecraft's spin-axis along the Sun-Earth line \citep{chiu1998ace, stone1998advanced}}. In all cases, the total marginal spectrum (black curve) exhibits a behaviour consistent with the energy cascade that dissipates below the proton gyrofrequency.

\subsection{Arbitrary order Hilbert Spectra}

Turbulent flows exhibit intermittency and scaling behaviours across various statistical moments and scales. For a generic field component $x(t)$, these properties are analysed through field fluctuations and defined as increments separated by time $\tau$ \citep{bruno2013solar},
\begin{equation}
    \Delta x_{\tau} = x(t + \tau) - x(t) \ .
\end{equation}
These fluctuations characterise the intermittency and multiscaling properties of the system and are quantified using a structure function of order $q \ (q>0)$, $S_{q}$. Mathematically, this is represented as,
\begin{equation}
    S_{q} \equiv \, \langle |\Delta x_{\tau}|^{q} \rangle \, \sim \, \tau^{\zeta(q)} \ ,
\end{equation}
where $\zeta(q)$ is the scaling exponent function of order $q$ that characterises these fluctuations. In a fully developed turbulence, structure functions describe the evolution of fluctuations across different scales. For monofractal systems, $\zeta({q})$ follows a linear trend, but deviates towards non-linear behaviour for multifractal systems, displaying the intermittent nature of turbulence \citep{bruno1998turbulence}.

\rev{To compute an arbitrary higher-order ($q>0$) marginal Hilbert spectrum $\mathcal{L}_{q}(\omega)$, \cite{huang2008amplitude} proposed a method based on the joint probability density function $p(\omega,a)$ of the instantaneous frequency $\omega$ and instantaneous amplitude $a$,}
\begin{equation}
    \mathcal{L}_{q}(\omega) = \int_{0}^{\infty} p(\omega, a)\, a^{q}\,da \ . \label{eq:lq}
\end{equation}
\rev{For an arbitrary order $q$, the power spectra follows the scaling relation, $\mathcal{L}_{q}(\omega) \sim \omega^{-\alpha_q}$, where $\alpha_{q}$ is the spectral slope. The scaling exponent $\zeta({q})$ can then be related to $\alpha_{q}$ computed from Eq. \ref{eq:lq} as,}
\begin{equation}
    \rev{\zeta(q) = \alpha_{q} - 1 \ .}
\end{equation}
\rev{The spectral exponents $\alpha_{q}$ can then be considered analogous to the $\zeta({q})$ exponents obtained from $S_{q}$ to identify intermittency in the signal
\citep{2011PhRvE..84a6208H, carbone2018arbitrary}.}

\subsubsection{Second order marginal Hilbert Spectra}

\begin{figure*}
    \centering
    \includegraphics[width=\linewidth]{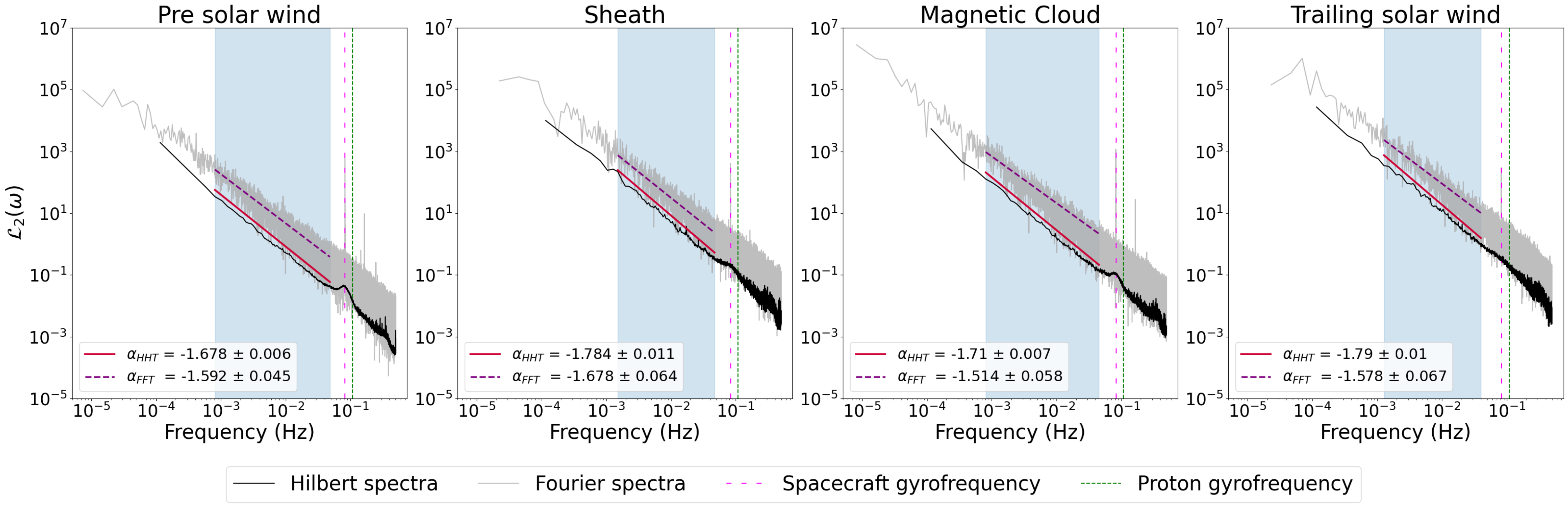}
    \caption{\rev{Second-order marginal Hilbert spectra $\mathcal{L}_{2}(\omega)$ (black curve) plotted against the Fourier spectra (grey curve) for the different ICME regions. The spectra were obtained by adding the second-order marginal Hilbert spectra of the components $B_{x}, B_{y}, B_{z}$ for each region. The selected inertial ranges are marked by the blue-shaded region. The best-fit lines for $\mathcal{L}_{2}(\omega)$ and Fourier PSD within this range are shown by the red line (vertically shifted for clear representation) and purple dashed line, respectively. The corresponding spectral indices are denoted by $\alpha_{HHT}$ and $\alpha_{FFT}$ in each subplot.
    The vertical magenta and green dashed lines mark the spacecraft gyrofrequency and proton gyrofrequency, respectively.}}
    \label{mhs_order_2}
\end{figure*}

For the second-order case, the marginal Hilbert spectrum takes the form,
\begin{equation}
    \mathcal{L}_{2}(\omega) = \int_{0}^{\infty} p(\omega, a)\, a^{2}\,da \ .\label{eq:l2}
\end{equation}
Eq. \ref{eq:l2} is comparable to the Fourier power spectral density, which provides energy distribution across different frequencies. Following this, second-order marginal Hilbert spectra were generated for each magnetic field component across the \rev{ICME regions. The marginal Hilbert spectra of the field components, $B_{x}$, $B_{y}$, and $B_{z}$ were added together to obtain the second-order marginal Hilbert spectrum of the complete region, as shown by the black curve in Fig. \ref{mhs_order_2}.} 

The plots show distinct scaling behaviours corresponding to the inertial and dissipation ranges. The inertial range is associated with larger time scales (lower frequencies), while the dissipation range corresponds to smaller time scales (higher frequencies), occurring beyond the proton gyrofrequency (indicated by the green dashed line). \rev{To analyse the statistical properties of our data, we selected a region within the inertial range using piece-wise linear fitting, as indicated by the blue-shaded region in Fig. \ref{mhs_order_2}. Based on the $\mathcal{L}_{2}(\omega)$ spectra, the inertial range was selected separately for each ICME region but was restricted within $\approx 8.1 \times 10^{-4}$ to $4.86 \times 10^{-2}$ Hz.} We confined our analysis within this range to exclude the influence of two known spectral artefacts: the spectral break at \rev{$\approx 3.4 \times 10^{-4}$} Hz and the spacecraft spin frequency $\omega_{sc}$, marked by the vertical magenta dashed line.

The gyrofrequency component of the spacecraft is identified in the second-order marginal Hilbert spectra as a distinct bump that peaks at $\omega_{sc}$. \rev{The effect of this feature appears to become less prominent with the increasing spectral power as the ICME propagates from the pre-shock to post-MC stages. This results in a flattening of the spectrum around $\approx \revii{4.86} \times 10^{-2}$ Hz, occurring just before the spectral bump observed at $\omega_{sc}$.} In the absence of this instrumental effect, a smoother spectral profile would be expected. However, removing or filtering out the gyrofrequency components associated with $\omega_{sc}$ is not a straightforward task, as these appear across different oscillatory modes, particularly in IMFs 2, 3 and 4 (Fig. \ref{marginal_amp}). Filtering out these frequency components while preserving the signal information requires specialised techniques, which are beyond the scope of this work, but may be addressed in future studies.

The spectral break observed at \rev{$\approx 3.4 \times 10^{-4}$} Hz arises from the binning scheme used to construct the Hilbert spectrum, where instantaneous frequencies were grouped into \rev{2160} equally spaced bins with a resolution of \rev{$\Delta f_{b} = 2.3 \times 10^{-4}$ Hz} (as discussed in \cref{subsec_hilbert_spectra}). \rev{The first bin contains all instantaneous frequencies between 0 and $\Delta f_{b}$ Hz and therefore exhibits the highest number of counts compared to the subsequent bins (Fig.~\ref{freq_dist}). Consequently, the spectral value in this bin reflects the cumulative background energy at low frequencies ($\omega_{j} < \Delta f_{b}$ Hz), resulting in a relatively elevated value.} Although increasing the bin size would provide finer frequency resolution and produce a more gradual spectral decay, such changes would have a negligible impact on the overall statistical properties of the data within the inertial range. Keeping the above two constraints in consideration, we restricted our analysis of the inertial range to frequencies between \rev{$\approx 8.1 \times 10^{-4}$ to $4.86 \times 10^{-2}$ Hz.}

The second-order marginal spectra for each ICME region presented in Fig. \ref{mhs_order_2} follow a scaling behaviour that can be characterised by a power law of the form,
\begin{equation}
    \mathcal{L}_{2}(\omega) = e^{c} \cdot \omega^{\alpha} \ ,
    \label{eq:power_law}
\end{equation}
where $c$ is a constant, and $\alpha$ is the spectral index. Applying the logarithmic function on both sides of Eq. \ref{eq:power_law}, we obtain,
\begin{equation}
    \log(\mathcal{L}_{2}(\omega)) = c \, + \alpha \log(\omega) \ ,
\end{equation}
which corresponds to a linear relationship. A linear regression was performed within \rev{the selected inertial ranges for each region} to estimate the spectral index $\alpha$ from the resulting line of best fit, as indicated in red in Fig. \ref{mhs_order_2}. The estimated values of $\alpha$ for each \rev{ICME region} are provided within the respective subplots.

\rev{Furthermore, we compared the $\mathcal{L}_{2}(\omega)$ with the standard Fourier power spectral density (PSD) for each ICME region, as depicted by the grey curve in each subplot of Fig. \ref{mhs_order_2}. The best-fit lines for $\mathcal{L}_{2}(\omega)$ and Fourier PSD in the selected inertial range for each ICME region are marked by the red solid line and purple dashed line, respectively, and their respective spectral indices are denoted as $\alpha_{HHT}$ and $\alpha_{FFT}$ in each subplot. From Fig. \ref{mhs_order_2}, we notice a better fit obtained for the $\mathcal{L}_{2}(\omega)$ spectra as compared to that of the Fourier PSD. The second-order marginal Hilbert spectrum is noticeably smoother than the Fourier PSD, which exhibits significant noise levels across the frequency range. While the energy at a given frequency in the Fourier PSD implies the presence of wave activity throughout the duration of the event, contrarily, the marginal Hilbert spectrum suggests the existence of the wave to have appeared locally, resulting in a cleaner spectral profile and allowing a clearer identification of the turbulent scaling behaviour within different ICME regions.}

The analysis of the marginal Hilbert spectra reveals clear variations in the inertial-range spectral slopes across the different ICME regions, indicating changes in the nature of the turbulent cascade throughout the event. The \rev{preceding solar wind} records an energy transfer in the inertial range with a spectral decay of \rev{$\alpha_{HHT} = -1.677 \pm 0.006$}. This value is equivalent to Kolmogorov's theoretically derived slope of -5/3, indicating the presence of a fully developed magnetohydrodynamic turbulence. The results are in agreement with previous studies that reported the presence of fully developed turbulence in the solar wind at 1 AU \citep{podesta2007spectral, telloni2021evolution}. 

In the \rev{sheath}, the marginal Hilbert spectrum within the inertial range exhibits a notably steeper slope, measured at \rev{$\alpha_{HHT} = -1.784 \pm 0.011$}, suggesting a more rapid energy decay at smaller scales. This feature is consistent with the typical low Alfvénic nature of the slow solar wind, where the turbulence energy is increased following the passage of the ICME shock. The shock-wave is capable of compressing and heating-up the solar wind plasma, leading to enhanced non-linear interactions, stronger intermittency, and anisotropic behaviour \rev{\citep{yordanova2009turbulence, bruno2013solar, teodorescu2015inertial, kilpua2021statistical, kilpua2021multi}}. 

\revii{The MC region reported a slower energy cascade following the sheath with $\alpha_{HHT} = -1.71 \pm 0.007$. This behaviour likely indicates that the mechanisms responsible for the spectral steepness in the sheath region are diminished in the large-scale flux ropes. A similar reduction in magnetic field spectral steepening in the MC, compared to the sheath, has also been reported in previous studies by \cite{sorriso2021turbulent, riazantseva2024linking}.} 

Lastly, the marginal Hilbert spectrum for the \rev{trailing solar wind region} reveals an even steeper slope of \rev{$\alpha_{HHT} = -1.79 \pm 0.01$, indicating the most rapid energy decay} observed during the event, particularly towards its termination. This steepening is likely associated with the large-scale magnetic field fluctuations observed in the later stages of the MC and the trailing solar wind, as illustrated in Fig. \ref{CME_B_field}. \rev{From the results of the Hilbert spectra (Fig. \ref{hilbert_spectrum}), the fluctuations exhibit enhanced amplitudes, particularly in the trailing solar wind region. This enhancement may be associated with the interaction of a fast solar wind stream with the ICME after June 30 00:00, thereby contributing to the increased power observed in this interval.}

\rev{Overall, our results obtained from the EMD-HSA framework provide a smoother energy spectra as compared to the standard Fourier PSD approach. The results also show that the turbulent energy cascade spectra changes with each region as the ICME progresses.}


\section{Conclusions}

In this work, we analysed the ICME event of 27 June 2013 \citep{prete2024euhforia} recorded by the MAG instrument onboard NASA's ACE spacecraft to study the turbulent behaviour at different stages of the ICME. The event was characterised by the onset of a shock which led to the formation of a sheath region with increased magnetic field activity, followed by an expanding magnetic cloud. The event was divided into four distinct regions --- the solar wind preceding the ICME shock, \revii{followed by the sheath, the MC and the trailing solar wind.}

To study ICME turbulence, the magnetic field data in the GSE coordinates for the components $B_{x}$, $B_{y}$, and $B_{z}$ across the \rev{four ICME regions} were analysed using the EMD-HSA methodology. The magnetic field signals were decomposed into simpler, finite intrinsic mode functions ordered from the highest to the lowest oscillatory frequency modes through empirical mode decomposition. The extracted IMFs were subjected to Hilbert spectral analysis to derive their corresponding instantaneous frequencies and amplitudes as functions of time. These quantities were used to generate the Hilbert spectrum for each magnetic field component within each region, providing a time-frequency distribution of the amplitude. The resulting spectrograms reported that the dominant amplitude contributions developed from the low instantaneous frequencies (< 0.1 Hz) present in the magnetic field components in the four regions. \rev{The Hilbert spectra for the sheath and MC reported sharp peaks, particularly in the $B_{y}$ and $B_{z}$ showing enhanced magnetic activity within these regions.} \rev{The Hilbert spectra further reported heightened magnetic field fluctuations in the three components of the trailing solar wind, following the termination of the MC, suggesting the significant after-effects of the ICME event persisting even after the cloud's passage, combined with a fast stream reported after June 30, 00:00:00.}

The second-order marginal Hilbert spectra $\mathcal{L}_{2}(\omega)$ were computed to map the power spectral density for each \rev{ICME region by combining the individual power spectral densities of the magnetic field components within the interval and a power-law fit was derived for each region within the selected inertial range.} The lower bound of the inertial range was selected to avoid energy contributions from frequencies below \rev{$\Delta f_{b} = 2.3 \times 10^{-4}$} Hz due to the adopted frequency-binning scheme, while the upper bound was chosen to exclude frequencies beyond $\revii{4.86} \times 10^{-2}$ Hz, affected by the gyration frequency of the spacecraft.

\rev{The marginal Hilbert spectral analysis reveals noticeable variations in the turbulent cascade across the ICME structure. The preceding solar wind exhibits a spectral slope close to the Kolmogorov value ($\alpha_{HHT}\approx-5/3$), indicating fully developed turbulence typical of ambient solar wind conditions at 1 AU. In contrast, the sheath and trailing solar wind regions show noticeably steeper spectra ($\alpha_{HHT}\sim -1.78$ to $-1.79$), suggesting enhanced intermittency and the increased presence of discontinuities and current sheets generated by shock compression and subsequent solar wind–ICME interactions. These findings are consistent with previous observations of ICME-driven turbulence reported in sheath regions. Within the magnetic cloud, the spectral slope is found to be less steep, possibly due to the influence of the large-scale flux-rope structure. Furthermore, the EMD-HSA spectral approach provides smoother and more stable estimates of inertial-range scaling compared with the conventional Fourier PSD, thus highlighting its capability to characterise turbulence in highly non-stationary solar wind structures such as ICMEs.}




\begin{acknowledgements}
We thank the anonymous referee for their constructive comments that improved the manuscript.
\\
We thank the ACE MAG instrument team and the ACE Science Center for providing the ACE data. The magnetic field data of the ICME analysed in this study was downloaded from NASA's Space Physics Data Facility (SPDF)\footnote{\url{https://spdf.gsfc.nasa.gov/}} through the Coordinated Data Analysis Web (CDAWeb)\footnote{\url{https://cdaweb.gsfc.nasa.gov/}} database system. The ACE Magnetic Field (MAG) Level 2 data that supports the findings of this study is made available by NASA SPDF \citep{Smith_Ness_2022}.
\\
This work made use of the \verb|Python| programming language and the following software packages for analysis: \textsc{Jupyter} \citep{granger2021jupyter} ; \textsc{Matplotlib} \citep{Hunter:2007} ; \textsc{Numpy} \citep{harris2020array} ; \textsc{Pandas} \citep{mckinney-proc-scipy-2010} ; \textsc{Piecewise Linear Fit} \citep{pwlf} ; \textsc{PyEMD} \citep{pyemd} ; \textsc{SciPy} \citep{2020SciPy-NMeth}.
\\
This publication was produced while attending the PhD program in PhD in Space Science and Technology at the University of Trento / University of Calabria, Cycle XXXIX, with the support of a scholarship financed by the Ministerial Decree no. 118 of 2nd March 2023, based on the NRRP - funded by the European Union - NextGenerationEU - Mission 4 "Education and Research", Component 1 "Enhancement of the offer of educational services: from nurseries to universities” - Investment 4.1 “Extension of the number of research doctorates and innovative doctorates for public administration and cultural heritage” - CUP [E66E23000110001].
\end{acknowledgements}

%

\bibliographystyle{aa} 
\bibliography{references} 
%

\end{document}